\documentclass[12pt]{article}
\usepackage[margin=2.5cm]{geometry}  
\usepackage{epsfig, graphics}

\usepackage{subcaption}

\usepackage{caption}

\usepackage{amsfonts}
\usepackage{amssymb}
\usepackage{amsmath}
\usepackage[latin1]{inputenc}

\usepackage{xcolor}
\usepackage{multirow}
\usepackage{hyperref}
\usepackage{amsmath}
\usepackage{graphicx}
\usepackage{tikz}
\usepackage{circuitikz}
\usetikzlibrary{arrows,shapes,positioning}
\usepackage[margin=2.5cm]{geometry}  
\usepackage{graphicx,psfrag,epsf}
\usepackage{enumerate}
\usepackage{url} 
\usepackage{siunitx}
\usepackage{float}
\usepackage{algorithm}
\usepackage{algorithmic}
\usepackage{amsmath}
\usepackage{amssymb}
\usepackage{graphicx}
\usepackage{subcaption}
\usepackage{natbib}
\usepackage{lineno}
\usepackage{dcolumn}
\usepackage{multirow}
\usepackage{comment}
\usepackage{booktabs}
\usepackage{colortbl}
\usepackage{rotating}
\usepackage{epstopdf}
\usepackage{xcolor}  
\usepackage{hyperref}
\hypersetup{
	    colorlinks=true,
    linkcolor=blue,
    urlcolor=blue,
    citecolor=blue
}
\usepackage{float}
\newcommand{\nin}{\noindent}

\def\bkR{{\rm I\kern-.17em R}}

\def \1n{1\hskip -3pt \mbox{N}}

\newfont{\bbf}{cmbx12 scaled 1435}

\usepackage[margin=2.5cm]{geometry}  
\usepackage{graphicx,psfrag,epsf}
\usepackage{enumerate}
\usepackage{url} 
\usepackage{siunitx}
\usepackage{float}
\usepackage{algorithm}
\usepackage{algorithmic}
\usepackage{amsmath}
\usepackage{amssymb}
\usepackage{graphicx}
\usepackage{subcaption}
\usepackage{natbib}
\usepackage{lineno}
\usepackage{dcolumn}
\usepackage{multirow}
\usepackage{comment}
\usepackage{booktabs}
\usepackage{colortbl}
\usepackage{rotating}
\usepackage{epstopdf}
\usepackage{xcolor}  
 \newcolumntype{d}[1]{D{.}{.}{#1}}
\usepackage{hyperref}
\hypersetup{
    colorlinks=true,
    linkcolor=blue,
    urlcolor=blue,
    citecolor=blue
    }

\usepackage{rotating}
\usepackage{epsfig}%
\usepackage{epstopdf}
\begin{document}
\setlength{\baselineskip}{.26in}
\thispagestyle{empty}
\renewcommand{\thefootnote}{\fnsymbol{footnote}}
\vspace*{0cm}
\begin{center}

\setlength{\baselineskip}{.32in}
{\bbf Shrinkage Regularization for (Non)Linear Serial Dependence Test }

\vspace{0.5in}

\large{Francesco Giancaterini}\footnote{Universit`a di Roma "Tor Vergata", Italy, e-mail:{\it francesco.giancaterini@uniroma2.it}}\footnote{Fondazione Ugo Bordoni, Rome, Italy, e-mail: {\it fgiancaterini@fub.it}}
 \large{Alain Hecq}\footnote{Maastricht University, The Netherlands, 
e-mail:{\it a.hecq@maastrichtuniversity.nl}}, \large{Joann Jasiak}\footnote{York University, Canada, 
e-mail:{\it jasiakj@yorku.ca}},\large{ Aryan Manafi Neyazi }\footnote{York University, Canada, e-mail: {\it aryanmn@yorku.ca } \\

The authors acknowledge the financial support of MITACS and NSERC Canada.
}

\setlength{\baselineskip}{.26in}

 \vspace{0.4in}

This version: \today\\

\medskip

\vspace{0.3in}
\begin{minipage}[t]{12cm}
\small
\begin{center}
{\bf Abstract} \\
\end{center}

This paper introduces a regularized test of the null hypothesis of the absence of linear and nonlinear serial dependence for high-dimensional non-Gaussian time series. Our approach extends the portmanteau test introduced in \cite{jasiakneyazi} to the high-dimensional setting.

\bigskip


\bigskip


\end{minipage}

\end{center}
\renewcommand{\thefootnote}{\arabic{footnote}}

\newpage
\section{Introduction}

The nonlinear serial dependence test (NLSD hereafter) introduced by \cite{jasiakneyazi} is a portmanteau-type test based on the autocovariances of nonlinear functions of strictly stationary time series with non-Gaussian distributions. This approach follows from \cite{chan2006note} and has been further developed by \cite{gourieroux2023generalized} for inference on mixed causal-noncausal models from the GCov estimator and the GCov-based specification test. In the frequency domain, a nonlinear serial dependence test was introduced by \cite{escanciano2009automatic}.

The NLSD test can be used to identify nonlinear and noncausal dynamics in both univariate and multivariate time series. The test statistic is computed from nonlinear transforms $a(.)$ of a univariate or multivariate non-Gaussian process $\{X_t,\}, t=1,...,T$ of dimension $N$. The transformed vector $X_t^a$ is of dimension higher than $N$, because it is augmented by $K$ nonlinear functions of $X_t$, such as the squares, absolute values or logarithms, for example, satisfying the regularity conditions given in \cite{gourieroux2023generalized}. For example, let us consider $K$ nonlinear functions $a_1,...,a_K$ of a univariate $X_t$ with $N=1$. These functions transform $X_t$ into a multivariate process of dimension $NK=K$ with components $a_k(X_t)$ as follows:

\begin{equation}
    X_t^a = \left( \begin{array}{c} a_1(X_{t}) \\
                          \vdots \\
                          a_K (X_t)
                          \end{array}\right),   
\end{equation}

\nin where $a_1(X_t) = X_t$ is the vector time series itself to capture the linear dependence. Note that if $X_t$ has no finite second-order moments, then it can be replaced by a transformed multivariate process $X_t^a$ with finite moments. Then, from a sample of $T$ observations on $X_t$, we can compute the sample autocovariance matrix of the transformations $a_k(X_t), k=1,...,K$ :
\begin{equation}
\label{Gamma}
    \hat{\Gamma}_T^a(h) = \frac{1}{T} \sum_{t=h+1}^T  X_t^a X_{t-h}^{a'} - \frac{1}{T} \sum_{t=1}^{T-h} X_t^{a} \frac{1}{T} \sum_{t=h+1}^T  X_{t-h}^{a'}. 
\end{equation}

\nin and the sample variance matrix $\hat{\Gamma}_T^a(0)$. Let us  consider the following set of null hypotheses: $H_{0,a} = (\Gamma^a (h) = 0, \; h=1,...,H).$ To test for the absence of linear and nonlinear serial dependence, we compute the test statistic:
\begin{equation}
\label{eq:1}
\hat{\xi}^a_T(H)= T \sum_{h=1}^H Tr \hat{R}_a^2(h),
\end{equation}
where
$$ \hat{R}_a^2(h)=\hat{\Gamma}_T^a(h) \hat{\Gamma}_T^a(0)^{-1}\hat{\Gamma}_T^a(h)' \hat{\Gamma}_T^a(0)^{-1}.$$

\nin Under the null hypothesis of independence, the NLSD test statistic $\hat{\xi}^a_T(H)$ follows asymptotically a $\chi^2((NK)^2 H)$ distribution \footnote{\cite{jasiakneyazi} show that for an increasing set of well-selected nonlinear transformations the null hypothesis of the absence of (non)linear dependence is close to the independence hypothesis, which is the implicit null hypothesis used for the derivation of the asymptotic properties of the test.}.

When $N$ or $K$ or both are large then the matrix $\hat{\Gamma}_T^a(0)$
is of high dimension and its inverse may be difficult to compute. Two approaches in the literature address this issue. The first one consists in replacing the sample variance matrix by a matrix containing only its diagonal elements, and was proposed by \cite{gourieroux2017noncausal}. The test statistics computed under this approach does not have an asymptotic chi-square distribution under the null of independence. The second is the Ridge regularization (RNLSD) test proposed by \cite{giancaterini2025regularized}, which relies on a Ridge-type regularization of the NLSD test and yields a RNLSD test statistic with an asymptotic chi-square distribution under the null of independence. \cite{giancaterini2025regularized} suggest the cross-validation for selecting the optimal regularization parameter. In this paper, we consider the shrinkage approach of \cite{ledoit2004well} for estimating the sample covariance matrix, when it suffers from the curse of dimensionality. The advantage of this approach is that we can estimate the shrinkage parameter in a single step, directly from the sample. The shrinkage Regularized NLSD (SR-NLSD hereafter) test has an asymptotic chi-square distribution with known degrees of freedom under the null of independence and relies on consistent estimators of the shrinkage parameters.

\bigskip

\section{ Linear Shrinkage Estimator of Ledoit and Wolf}

\cite{ledoit2004well} propose a general asymptotic framework, which is reviewed below using their generic notation and allows  $p=N$ to tend to infinity at a rate such that $p/T$ stays bounded. They consider $X$, which is a $p \times T$ matrix of $T$ i.i.d observations on random variables of dimension $p$ with mean zero and variance matrix $\Sigma$. The Frobenius norm is defined as $||A||=\sqrt{tr(AA')/p}$, based on the quadratic form of $||.||^2$ and the inner product $<A_1,A_2>=tr(A_1A_2')/p$. The sample covariance matrix is denoted by $S=XX'/T$. The goal is to find the linear combination of identity matrix $I$ and sample covariance matrix $S$, which are the components of a regularized variance estimator $\Sigma^*= \rho_1 I + \rho_2 S$ which minimize $E [||\Sigma^* - \Sigma||^2]$, the expected distance between the regularized variance estimator and true variance and $\rho_1, \rho_2$ are the tuning parameters. \cite{ledoit2004well} identify four scalars that play important roles:

$$\mu=<\Sigma, I>, \;\; \alpha^2=||\Sigma-\mu I||^2, \;\;\beta^2 = E[||S-I||^2] , \;\; \delta^2=E[||S-\mu I||^2],$$ 

\noindent where $||.||$ denotes the Frobenius norm. For $\beta^2$ and $\delta^2$ to be finite we need to assume $X$ has a finite fourth moment.  Theorem 2.1 of \cite{ledoit2004well} shows that the solution to the optimization problem of 
\begin{equation}
\label{opt}
\begin{aligned}
\underset{\rho_1,\varphi_2}{Min} \quad &  E[||\Sigma^*-\Sigma||^2]   \\    
  \textrm{s.t.} \quad & \Sigma^*= \rho_1 I + \rho_2 S \\ 
  \end{aligned}
\end{equation}

\noindent where $\rho_1$ and $\rho_2$ are non-random regularization parameters  is

\begin{equation}
       \label{solution}
       \Sigma^*= \frac{\beta^2}{\delta^2} \mu I + \frac{\alpha^2}{\delta^2} S ,
\end{equation}
\noindent and the objective function is:

\begin{equation}
       E[||\Sigma^*-\Sigma||^2]=\frac{\alpha^2 \beta^2}{\delta^2}.
\end{equation}

\nin \cite{ledoit2004well} propose a consistent estimator of $\Sigma^*$, which is independent of the population variance $\Sigma$ and based on a sample of $T$ observations on i.i.d. variables of dimension $p$ arranged in matrix $X_T$. They consider the spectral decomposition of the variance matrix into eigenvectors and eigenvalues: $\Sigma_T = G_T \Lambda_T G_T '$ where $G_T$ is the rotation matrix whose each column is the eigenvectors of $\Sigma_T$ and $\Lambda_T$ is a diagonal matrix whose elements are the eigenvalues of $\Sigma_T$. Consider $Y_T=G'_T X_T$ and let $(y^T_{11},...,y^T_{p 1})'$ be the first column of $Y_T$. 

\nin To define the  consistent estimator of $\Sigma^*$, we need the following assumptions:

\nin \textbf{Assumption 1.} There exists a constant $\kappa_2$ independent of the number of observations such that
$$  \frac{1}{p} \sum_{i=1}^{p}  E[(y^T_{i1})^8] \leq \kappa_2.$$

\medskip
\nin \textbf{Assumption 2.}

$$\lim_{T \to  \infty}\frac{(p)^2}{T^2} \times \frac{\sum_{(i,j,k,l)\in Q_T}  (Cov[y^T_{i1}y^T_{j1},y^T_{k1}y^T_{l1}])^2}{Card(Q_T)} =0,$$
Where $Q_T$ is a set of all quadruples of distinct integers of 1 to $p$ and $Card(Q_T)$ gives the number of elements in the set $Q_T$. 

\medskip
\nin \textbf{Theorem 3.2, \cite{ledoit2004well}:} 

\nin Suppose there exists a constant $\kappa_1 \in [0,\infty)$ independent of the number of observations $T$ such that $\frac{p}{T}\to \kappa_1$. Then, under Assumptions 1 and 2, the estimator 

\begin{equation}
    S_T^*=\frac{b^2_T}{d^2_T} m_T I+ \frac{a^2_T}{d^2_T} S_T,
\end{equation}
is a consistent estimator of $\Sigma_T^*$,where:
$$m_T=<S_T,I> , d^2_T=||S_T-m_T I||^2, a^2_T=d^2_T-b^2_T ,b^2_T= min(d^2_T,\Bar{b}^2_T),$$
and:
$$\Bar{b}^2_T=\frac{1}{T^2} \sum_{i=1}^T || X_i X_i' - S_T||^2 $$ 
where $X_i $ is the $ith$ column of $X_T$ of dimension  $p\times 1$. 

\medskip

\nin Note that the proposed estimator $S^*_T$ is a consistent estimator of $\Sigma^*_T$ and not of $\Sigma$. From the above theorem, it follows that  $S^*_T$ is consistent estimator of $\Sigma$ if $$\frac{p}{T} \to 0$$

\nin Then, $m^2_T+ \text{Var}^2\left( \frac{1}{p} \sum_{i=1}^{p} (y_{i1})^2 \right) \to 0,$ and the tunning parameter estimators $\hat{\rho}_{1,T} = \frac{b^2_T}{d^2_T} m_T $ and $\hat{\rho}_{2,T} = \frac{a^2_T}{d^2_T} $
tend to deterministic limits $\hat{\rho}_{1,T} \rightarrow \rho_1=0$
and $\hat{\rho}_{1,T} \rightarrow \rho_1=1$ [\cite{pourahmadi2013high} Corollary 4.1].

\section{Shrinkage Regularization of the NLSD Test }

Under the null hypothesis of independence, the framework of \cite{ledoit2004well} can be applied to regularize the NLSD test statistic. 

Let us assume that all components of the multivariate process $\{ X_t^a\}, t=1,...T$ have mean zero or are demeaned. Then, the consecutive observations can be
arranged into a matrix  $X^a_T$ of dimension $p \times T$ where $p=NK$, which replaces matrix $X_T$ from Section 2. 
Then, under the null hypothesis of independence, $X^a_T$ contains i.i.d observations of dimension $p$ on random variables with mean zero and population variance matrix $\Gamma^a(0)$.

\nin Following the approach of \cite{ledoit2004well} and their their generic notation
we define: $${\Gamma^{a^*}(0)}=\rho_1 I+\rho_2 {\Gamma^{a}(0)}.$$

\nin Then, based on a sample of size T, we obtain a consistent estimator of ${\Gamma^{a^*}(0)}$ and of the tuning parameters:  

\begin{equation}
    {\hat{\Gamma}^{a^*}_T(0)}= \hat{\rho}_{1,T}  I + \hat{\rho}_{2,T}  \hat{\Gamma}^{a}_T(0) =  \frac{b^2_T}{d^2_T} m_T I+ \frac{a^2_T}{d^2_T} \hat{\Gamma}^{a}_T(0),
\end{equation}
where:
$$m_T=<\hat{\Gamma}^{a}_T(0),I> , d^2_T=||\hat{\Gamma}_T^{a}(0)-m_T I||^2, a^2_T=d^2_T-b^2_T ,b^2_T= min(d^2_T,\bar{b}^2_T),$$
and:
$$\Bar{b}^2_T=\frac{1}{T^2} \sum_{i=1}^T || X^a_i X^{a'}_i - \hat{\Gamma}_T^{a}(0)||^2 $$ 
with $X_i $ being a $p\times 1$ vector.

\medskip
\textbf{Definition 1:} The shrinkage-regularized NLSD test statistics (SR-NLSD) is
\begin{equation}
\label{eq:2}
\hat{\xi}^a_{SR}(H)= T \sum_{h=1}^H Tr \hat{R}_{SR}^2(h),
\end{equation}

\nin where $$ \hat{R}_{SR} ^2(h)=\hat{\Gamma}_T^a(h) {\hat{\Gamma}_T^{a^*}(0)}^{-1}\hat{\Gamma}_T^a(h)' {\hat{\Gamma}_T^{a^*}(0)}^{-1}$$

\medskip

\nin Recall that Theorem 3.2 of \cite{ledoit2004well} implies that  $\hat{\Gamma}^{a^*}_T(0)$ is consistent estimator of $\Gamma(0)$ if $\frac{p}{T} \to 0,$ which includes the case of $p=NK$ large and constant and $T \rightarrow \infty$, considered in \cite{giancaterini2025regularized}. In that context,
the autocovariance estimator maintains its standard asymptotic properties established by e.g. \cite{chitturi1976distribution}.
Then, the tunning parameter estimators $\hat{\rho}_{1,T} = \frac{b^2_T}{d^2_T} m_T $ and $\hat{\rho}_{2,T} = \frac{a^2_T}{d^2_T} $  defined above 
tend to deterministic limits $\hat{\rho}_{1,T} \rightarrow \rho_1=0$
and $\hat{\rho}_{2,T} \rightarrow \rho_2=1$

Under the Assumptions 1, 2 of \cite{ledoit2004well}, Assumption A.2 of \cite{giancaterini2025regularized}, we have the following result:

\nin \textbf{Proposition 1:}

If Assumptions 1-2 and A.2 are satisfied and if $\hat{\rho}_{1,T} \rightarrow 0$
and $\hat{\rho}_{2,T} \rightarrow 1$ when $p$ is large and $T \rightarrow \infty$ so that
$\frac{p}{T} \to 0$, then under the null hypothesis of independence,
 the SR-NLSD test statistic $\hat{\xi}^a_{SR}(H)$ has an asymptotic chi-square distribution with a degree of freedom equal to $p^2H$.

\medskip

\nin \textbf{Proof:} Based on  Theorem 3.2 of \cite{ledoit2004well} we know that ${\hat{\Gamma}_T^{a^*}(0)}$ is a consistent estimator of $\Gamma^{a^*}(0)=\rho_1I+\rho_2{\Gamma^{a}(0)}$ which itself is a consistent estimator of $\Gamma^{a}(0)$ when $\frac{p}{T} \to 0$. Therefore, the asymptotic distribution follows from the asymptotic equivalence of the SR-NLSD and NLSD test statistics.

The test consists in rejecting the null hypothesis of the absence of linear and nonlinear dependence if $\hat{\xi}^a_{SR}(H) > \chi^2_{0.95}(p^2H)$.

\section{Simulation Studies}

We investigate the empirical size of the NLSD and SR-NLSD tests. We consider two experiments. First, we generate i.i.d observations $X_t$ for $t=1,..T$, where $T=100,200,...,1000$ and $N=2,4,...,20$ from a Student's \textit{t}-distribution with degrees of freedom equal to 4, 7 and 10 and with variance equal to $I_N$. We consider $H=1$ and $K=2$ (level and power of the time series). Each square of Figure 1 plots is obtained by 1000 replications in the Monte Carlo, and in each experiment, we test the absence of linear or nonlinear dependence in the time series. We report the rate at which we reject the null hypothesis in both cases. In the second experiment, we use the same setup, but instead of increasing the number of variables, we increase the number of transformations as $K=2,4,...,20$, which includes the powers of time series up to the $K$ and fixed the number of variables $N=2$. Figure 2 illustrates the empirical size of the NLSD and SR-NLSD tests for the second experiment.

Figures 1 and 2 show that NLSD performs poorly in terms of empirical size in a high-dimensional setting with many variables or transformations. However, SR-NLSD provides an empirical size close to the nominal size. Comparing the SR-NLSD in many variables and many transformations, it is relatively more conservative in the experiment with many transformations.

\begin{figure}[H]
    \centering 
    \includegraphics[width=18cm, height = 20cm]{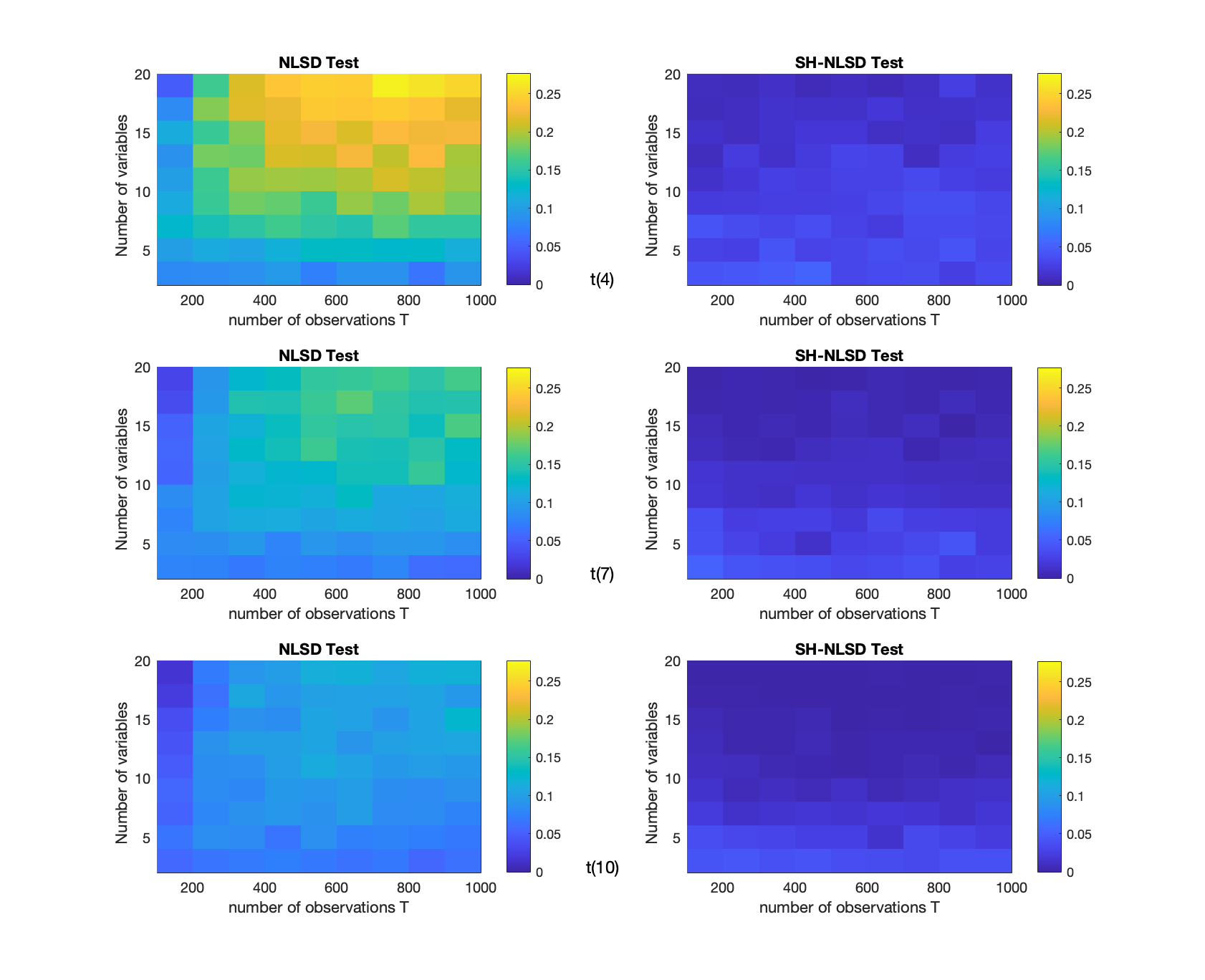}
    \caption{Empirical Size of NLSD test and SR-NLSD test for many variables N}
    \label{renixx}
\end{figure} 

\begin{figure}[H]
    \centering 
    \includegraphics[width=18cm, height = 20cm]{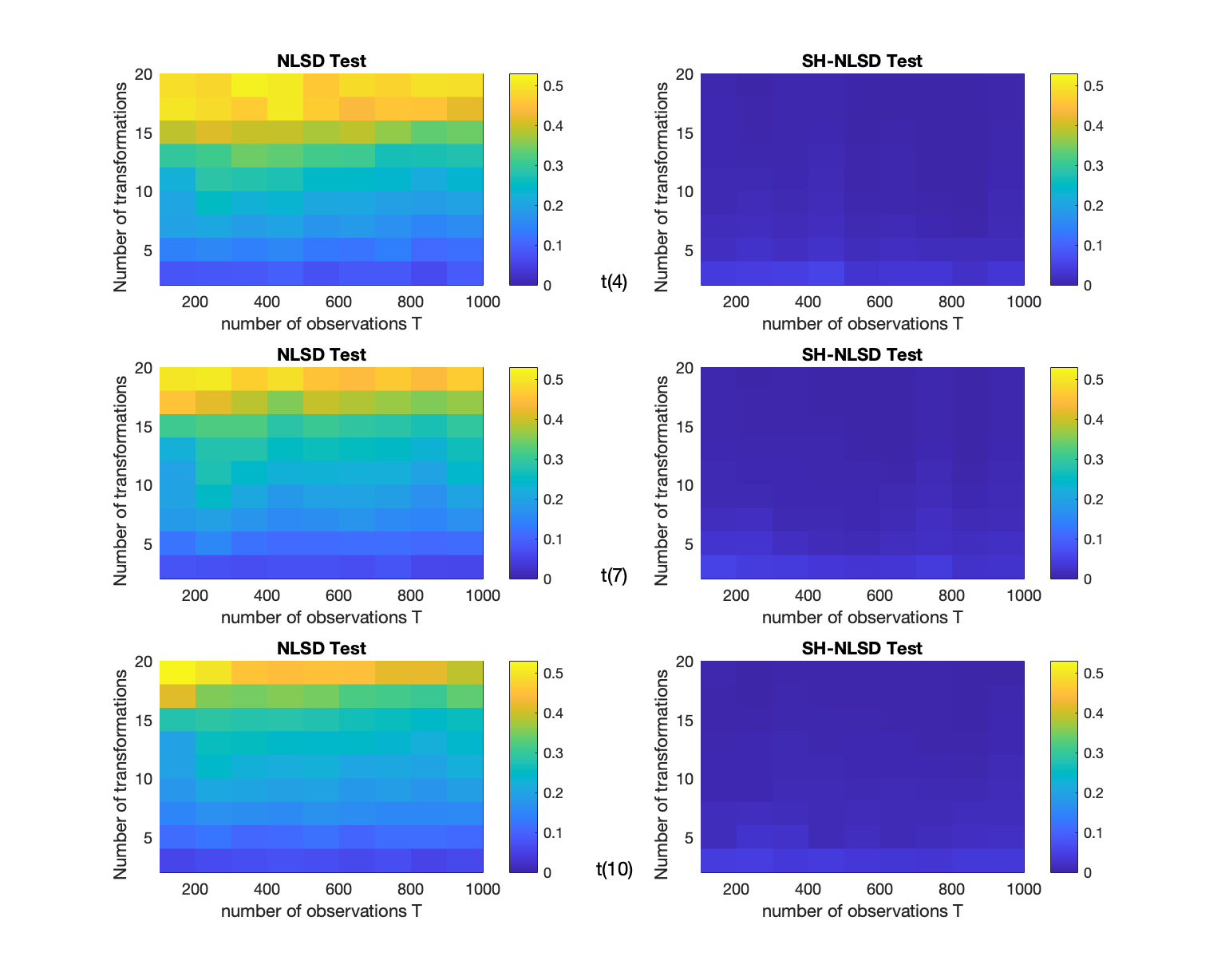}
    \caption{Empirical Size of NLSD test and SR-NLSD test, many transformations K}
    \label{renixx}
\end{figure}

\bibliographystyle{chicago}
\bibliography{refrences.bib}

\end{document}